\documentclass[12pt]{iopart}
\usepackage{graphicx}
\usepackage{gensymb}
\usepackage{color}
\usepackage{bm}

\begin{document}

\title[Correlated random walk of human embryonic stem cells \textit{in vitro}]{Correlated random walks of human embryonic stem cells \textit{in-vitro}}

\author{L E Wadkin$^1$, S Orozco-Fuentes$^1$, I Neganova$^2$, G Swan$^1$, A Laude$^3$, M Lako$^2$, A Shukurov$^1$ and N G Parker$^1$}
\address{$^1$ School of Mathematics, Statistics and Physics, Newcastle University, Newcastle upon Tyne, UK. }
\address{$^2$ Institute of Genetic
Medicine, Newcastle University, Newcastle upon Tyne, UK}
\address{$^3$ Bio-Imaging Unit, Medical School, Newcastle University, Newcastle upon Tyne, UK}
\ead{nick.parker@ncl.ac.uk}

\begin{abstract}
We perform a detailed analysis of the migratory motion of human embryonic stem cells in two-dimensions, both when isolated and in close proximity to another cell, recorded with time-lapse microscopic imaging. We show that isolated cells tend to perform an unusual locally anisotropic walk, moving backwards and forwards along a preferred local direction correlated over a timescale of around 50 minutes and aligned with the axis of the cell elongation. Increasing elongation of the cell shape is associated with increased instantaneous migration speed. We also show that two cells in close proximity tend to move in the same direction, with the average separation of $70\,\mu$m or less and the correlation length of around 25$\,\mu$m, a typical cell diameter. These results can be used as a basis for the mathematical modelling of the formation of clonal hESC colonies.
\end{abstract}


\submitto{\PB}
\maketitle

\section{Introduction}

There are many different types of active and spontaneous cell motion, e.g., swimming, gliding, crawling and swarming, detected in both prokaryotic and eukaryotic cells \cite{prokaryotes, eukaryoticreview}. The favour of one mechanism over another depends on the environment and the balance of achieved displacement and energy expenditure. Cell motility and migration is essential in many biological processes including the development, morphogenesis and regeneration of multicellular organisms, wound healing, tissue repair and angiogenesis \cite{Scarpa143,kurosaka08,AMAN201020,FRANZ2002153,li13,Lamalice782}. Anomalous cell migration can cause developmental abnormalities, tumour growth, neuronal migration disorders and the progression of metastatic cancer \cite{developmentabnormalities,Friedl03,Friedl12}. 

Unconstrained cell migration on a plane \textit{in vitro} can often be described as a two-dimensional random walk \cite{Codling}. The simplest random walk, the Brownian motion, is uncorrelated (the current direction of movement is independent of the last) and unbiased (the direction of each step is random). Correlated random walks (CRWs) involve a directional bias; there is a preference for the direction of the next step to be related to that in the previous step. It is common for cells in the absence of external biases to migrate as CRWs: the migration of amoeboids \cite{CRWamoeboid}, mammary epithelial cells \cite{CRWepithelial} and mouse fibroblasts \cite{CRWfibroblasts} have all been modelled as CRWs.  

Adaptations in cell morphology facilitate migration. Some eukaryotic cells achieve motion through the coordinated and cyclic reorganisation of the actin cytoskeleton, which determines their speed, direction and trajectory \cite{FilopodiaTentacles}. Several types of protrusive pseudopodia structures have been characterised, which mainly differ in the organisation of actin \cite{molecularbio}. Analysis of the formation of pseudopods has shown that cells extending pseudopodia which then split into two to allow a change of direction exhibit strong persistence and small turning angles \cite{CRWpseudopodia, CRWpseudopodia2}.

Understanding the form of cell trajectories provides important insights into diverse cell motility modes and helps to design and interpret experiments. For example, understanding the role of cell migration in metastatic cancer has led to new treatments which modify signalling pathways and alter cell morphology to reduce cell motility \cite{cancer1, cancer2}. A thorough understanding of the mechanisms underlying cell migration will not only deepen our understanding of many integral biological processes but also facilitate the development of therapies for treating migration-related disorders.

In this work we analyse the migration of human embryonic stem cells (hESCs) \textit{in vitro} on a homogenous two-dimensional matrix. Due to the promises of clinical applications of hESCs and hiPSCs (human induced pluripotent stem cells) and the discovery of new engineered substrates for cell growth, data presented in this paper is of prime importance \cite{LamininE8, NDND,SCtherapy}. Surface-engineered substrates provide an attractive cell culture platform for the production of clinically relevant factor-free reprogrammed cells from patient tissue samples and facilitate the definition of standardised scale-up methods for disease modelling and cell therapeutic applications \cite{Saha15112011}. Because the clinical application of stem cells may require as many as $10^10$ cells per patient and disease modelling efforts typically require more than $10^6$ cells to make a single differentiated cell type, robust methods of producing cells under conditions that accelerate proliferation could be particularly valuable \cite{pmid21960724,pmid21757228}. Feeder-free systems represent key progress in simplifying hESC/hiPSC production but most of these systems (synthetic polymers, peptide-modified surfaces, embryonic extra-cellular matrix (ECM) laminin isoforms, fibronectin from ECM with a small molecule mixture, and various vitronectin proteins) provide only modest gains in scaling-up hESC/hiPSC production because they still require seeding at a suitably high cell density and passaging through multicellular clumps. Even for the defined systems that support clonal growth \cite{pmid21478862,pmid21266967, pmid20729850}, mass production of synthetic polymers, recombinant proteins, or small molecule mixtures may be a challenge, particularly when considering the number of cells needed for disease modelling and clinical application. These matters highlight the need to understand the factors that may facilitate clonal expansion of hESCs and hiPSCs for clinical needs. Current efforts are focused on optimising differentiation protocols in order to generate homogenous populations of cells of interest, hence an understanding of the features that charcaterise the starting cell population is essential for informing these protocols.

Unfortunately, the motion and dynamics of single and pairs of hESCs/hiPSCs has received limited attention. In culture, hESCs are anchorage-dependent and migrate through actin cytoskeleton reorganisation \cite{adherentcells}. The main structures that define the leading edge on a migrating hESC are referred to as pseuodpodia. Motility is an intrinsic property of hESCs, and they perform an unbiased random walk when they are farther than $150\,\mu$m apart with cells closer to one another exhibiting coordinated motion \cite{Li}. Our previous work \cite{me} investigated how the kinematics of single and pairs of hESCs impact colony formation. We performed statistical analysis on cell mobility characteristics (speed, directionality, distance travelled and diffusivity) from the time-lapse imaging. We demonstrated that single and pairs of hESCs migrate as a diffusive random walk for at least 7 hours of evolution. We showed that for the cell pairs mutual interactions of closely positioned cells strongly affect the migration, and we identify two distinct behavioural regimes for cells resulting from a division. Also, the cell pair as a whole is shown to undergo a random walk with characteristic diffusivity \cite{me}. 

Here we focus on the migration of single and pairs of hESCs by examining more subtle and yet significant aspects of migration as a further step towards understanding cell group formation from a single cell. We consider how the direction of the motion is related to the cell morphology and analyse how the separation of cells affects their coordinated movements.

\section{Methods}
We follow the methods used to prepare and plate, and then image and track hESCs described in our previous work \cite{me}.  In brief, hESCs (WiCell, Madison WI) were plated at a density of 1500 cells/cm$^2$ onto 6-well plates pre-coated with Matrigel\textsuperscript{\textregistered} Basement Membrane Matrix (Corning Inc.), in the presence of mTeSR\textsuperscript{TM}1 media (STEMCELL Technologies). ROCKi (10$\mu$M, Chemdea) was present for the first hours after plating, and removed before time-lapse imaging.
  
After 1 hour, the plates were imaged with time-lapse microscopy (Nikon Eclipse Ti-E microscope) images taken every 15 minutes over 66 hours at a resolution of 0.62\,$\mu$m/pixel. From these images, we selected 26 single hESCs and 50 pairs of hESCs.  Single (isolated) hESCs are defined as those that initially have no neighbour within a 150$\,\mu$m radius; interactions of hESCs are negligible beyond this distance \cite{Li}.  The lineage trees for these cells are provided in Ref. \cite{me}.  We define the time variable $t$ as zero at the start of the image recording. The pairs of hESCs are those where the separation of two cells is less than $150\,\mu$m from each other and more than that from other cells. The cells either exist as pairs at the start of the imaging, or form a pair when a single isolated cell divides.

Each cell in our analysis was manually tracked throughout its motion, and its position in each image frame was defined as the location of its geometrical centre by eye, or `centre of mass' if the mass within the cell density is considered constant. For the single cell considered in Section 3.2, the cell boundary and geometrical centre was tracked using ImageJ \cite{fiji, imageJ2}. Comparison of this to the previous coordinates taken by eye showed no significant difference. Tracking of a single cell ceased when the cell died; cell pairs were tracked until one of them died or divided. We did not follow cell triples even when they were formed by division of a cell in a pair. Formation of a pair from convergence of two unrelated cells is rare since the individual random walks lead, on average, to the divergence of cell trajectories provided sufficient space is available. 

The instantaneous velocity of a cell was obtained from its displacement between two consecutive frames. Circular statistics calculations were performed as described in Ref. \cite{circstats} using Matlab and its Circular Statistics Toolbox (Directional Statistics) \cite{toolbox}.

\section{Results}

\begin{figure}[h]
\centering
\includegraphics[width=\textwidth]{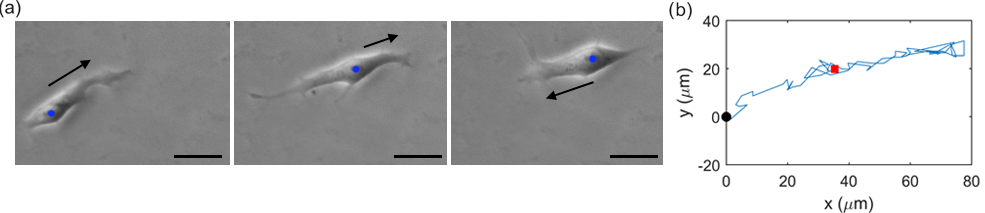} 
\caption{\label{fig:fb} (a) Images of a migrating single hESC. The frames are taken at $t=15\,$min, 6$\,$h$\,$45$\,$min and 14$\,$h$\,$15$\,$min. The blue dot shows the cell nucleus and the black arrow the direction of instantaneous velocity. The scale bars are 30$\,\mu$m in length. (b) Trajectory of the cell with the initial position (black dot) and final position (red square) shown.} 
\end{figure}

Figure \ref{fig:fb}(a) shows images of one of the cells during its migration; its full trajectory is shown in Figure \ref{fig:fb}(b). This cell is elongated in the instantaneous direction of motion, with a pseudopodia protrusion leading its next movement. The relation between motion and morphology is discussed in Section \ref{sec:morph}. The single cell shape can vary between approximately circular, with diameter of around 20$\,\mu$m, to more elongated with length of up to 70$\,\mu$m. In comparison, hESCs in colonies tend to be circular and considerably smaller, with diameters typically about $10\,\mu$m \cite{Li, Phadnis15}.

The cells can, and often do, change their direction of motion by up to $\pi$.  An example is shown in Figure~\ref{fig:fb}(a). The cell moves in the direction of its persistent pseudopodia protrusion, before contracting and moving in the direction of a new pseudopodia, resulting in a change of direction by approximately $\pi$. The whole manoeuver in this example takes about 6 hours. 

The lineage trees for the 26 single cells can be found in Ref. \cite{me}. Death rates are low, with only two cells dying before dividing. The remaining cells have divided by $t=20\,$h, with division occurring at a mean time interval of $t_{\scriptsize \textrm{d}}=7\,$h. The median speed of the cells is $16\,\mu$m/h, with the average of 23$\,\mu$m/h and with no noticeable differences in the migration behaviour between single cells which eventually die or divide.

\subsection{Single cells: correlated random walk}
\begin{figure}[b]
\centering
\includegraphics[width=\textwidth]{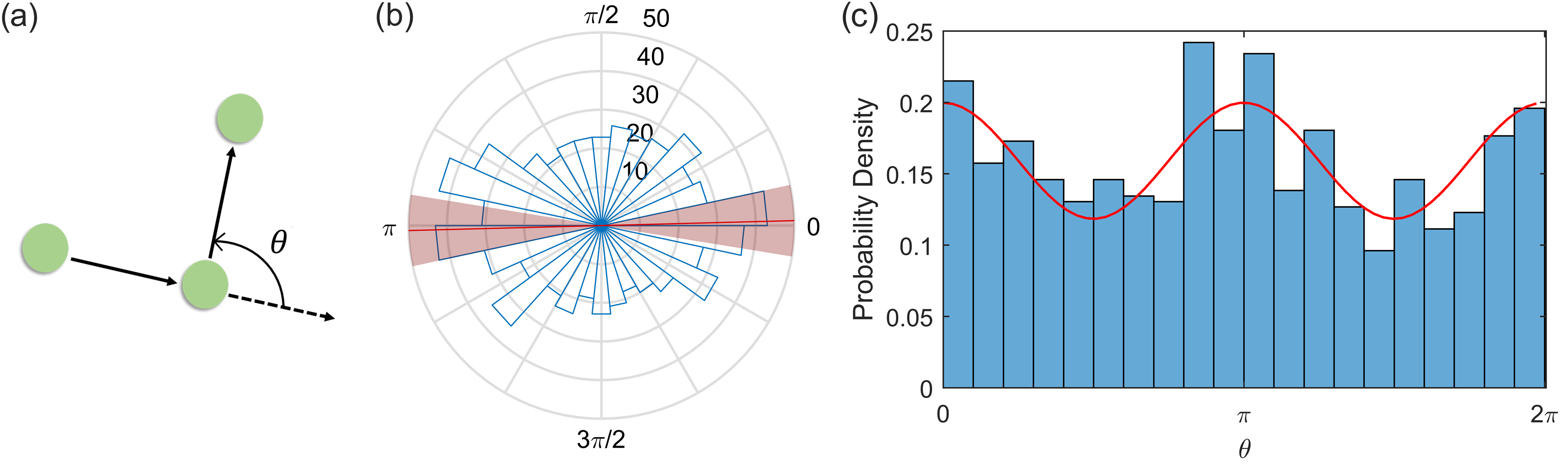} 
\caption{\label{fig:theta}(a) The definition of the turning angle $\theta$, the change in the cell's direction of motion from one time frame to the next. Green dots illustrate the positions of the cell in consequtive images with arrows representing the displacement vectors. (b) Polar histogram of $\theta$ for 26 single cells, over 18 hours, with 30 angular bins and 829 measurements. Overlaid the mean value of $0.026$ (red line) and one standard deviation ($0.19$) obtained by mapping the data to the range $0\leq\theta\leq\pi$ (pink shaded region). (c) The probability density of $\theta$ binned into 20 intervals. The least-squares fit $0.16+0.04\cos(2\theta)$ is shown in red.}
\end{figure}

First we seek to test for a bias in the direction of the single cell movements. We measured the turning angle, that is, the change in direction of the cell from one time frame to the next, denoted $\theta$ and illustrated in Figure~\ref{fig:theta}(a). As well as the turning angle with respect to the earlier direction of motion, we also considered the angle $\phi$ between the cell displacement and the global frame that does not change with time.

Figure ~\ref{fig:theta}(b) shows the polar histogram of $\theta$ for 26 single cells, while Figure~\ref{fig:theta}(c) presents the corresponding linear histogram. It is evident that the distribution has maxima at $\theta=0$ and $\theta=\pi$: the cell preferentially moves directly forwards or directly backwards with a roughly equal frequency between the two directions. The bias is robust, remaining even if small steps ($< 7 \mu$m) are removed from the dataset. The mean axis of movement, shown in Figure~\ref{fig:theta}(b), is approximately along the $\theta=0$ or $\theta=\pi$ (with the standard deviation of $\sigma_{\theta}=0.19$). In this manner, the motion represents a quasi-one-dimensional random walk. Both the $\chi^2$ and V tests reject the null hypothesis that the probability density of the turning angle $\theta$ is uniform at the 99.5\% confidence level. The probability density distribution can be approximated by $p=a+b \cos (2\theta)$ with $a=0.16$, $b=0.04$ and the $R^2$ value of 0.55. This fit suggests a symmetric spread of the distribution about $\theta=0$ and $\theta=\pi$.

\begin{figure}[b]
\centering
\includegraphics[width=\textwidth]{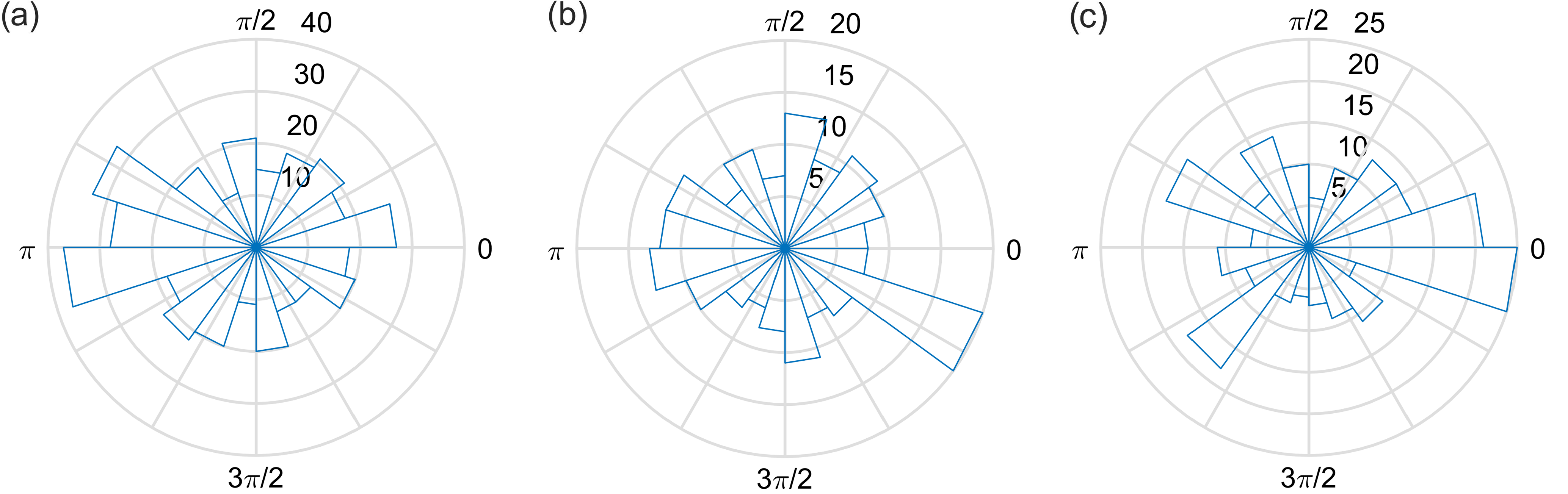} 
\caption{\label{fig:thetatime}Polar histograms for $\theta$ for all 26 single cells in the time intervals (a) 0--5\,h (26--12 cells, 20 bins and 404 measurements), (b) 5--10\,h (12--8 cells, 20 bins and 197 measurements) and (c) 10--18\,h (8--3 cells, 20 bins and 228 measurements).  There are fewer cells at later times due to cell divisions and deaths.} 
\end{figure}

\begin{figure}
\centering
\includegraphics[width=1\textwidth]{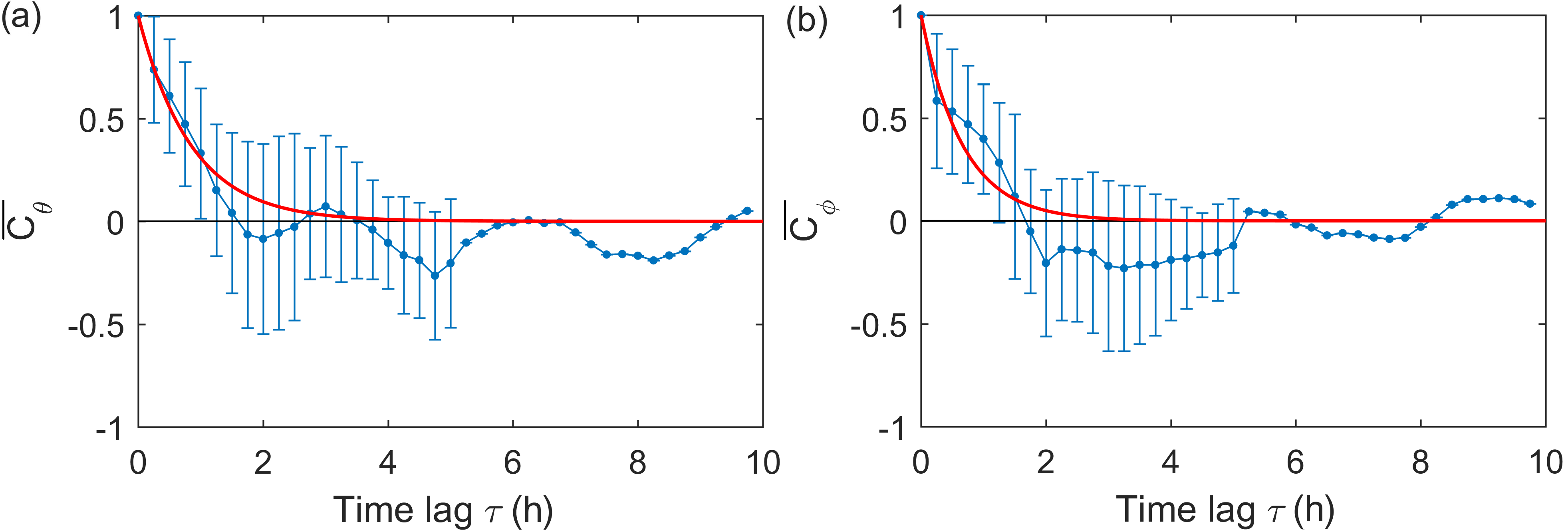} 
\caption{Average autocorrelation for (a) $\overline{C_{\theta}(\tau)}$ of $\theta$ and (b) $\overline{C_{\phi}(\tau)}$ of $\phi$ for single cells, with standard deviation error bars. The least squares fit (red line) is $\overline{C(\tau)}=e^{\tau/\tau_c}$ with (a) $\tau_{c}=0.8\pm0.1\,$h and (b) $\tau_{c}=0.7\pm0.2\,$h for $\theta$ and $\phi$, respectively. Cells are included in the average up to a lag of $N/3$. Note that onwards from a time lag of 5\,h, there is only one cell observed, hence the lack of error bars. Each lag corresponds to a time frame (15\,min). \label{fig:autocorr}} 
\end{figure}

The distribution of the turning angles has a distinct temporal pattern. Figure~\ref{fig:thetatime} shows the polar histograms of $\theta$ at early (0--5\,h), intermediate (5--10\,h) and late times (10--18\,h). At early times, the distribution is slightly biased towards $\theta=\pi$, indicating a weak dominance of the back-and-forth motion over a systematic forward motion. However, this effect is weak and the distribution is approximately uniform over angles. This is consistent with our previous observations that the motion of hESCs is close to an isotropic random walk at early times \cite{me}. By late times, however, the distribution is strongly biased towards $\theta=0$, that is, persistent forward motion. What we see on average for all times is a mixture of persistent and back-and-forth motions. This feature can be characterised with the temporal autocorrelation function, $C_{\theta}(\tau)$, for two-hourly moving averages of the angle $\theta$. For each cell, $C_{\theta}(\tau)$ is calculated as the circular correlation for $\theta$ with itself, delayed by a time lag of $\tau$. The average autocorrelation over all single cells, $\overline{C_{\theta}(\tau)}$, with least-squares fitting $\overline{C_{\theta}(\tau)}=e^{-\tau/\tau_c}$, $\tau_c=0.8$, is shown in Figure~\ref{fig:autocorr}(a). We see a temporal correation in $\theta$, with an average correlation time of $\tau_c=0.8\,$h.
 
We find no significant correlation in the global direction of movement, $\phi$, when individual steps are considered. We attribute this to the dominance of the back-and-forth motion over short periods of time. However, considering two-hourly moving averages we again find a systematic trend. The average autocorrelation over all single cells, $\overline{C_{\phi}(\tau)}$, is shown in Figure~\ref{fig:autocorr}(b). We see a temporal correlation in $\phi$, with least-squares fitting $\overline{C_{\phi}(\tau)}=e^{\tau/\tau_c}$, $\tau_c=0.7$ shown in Figure~\ref{fig:autocorr} and hence a correlation time of $\tau_c=0.7\,$h, similar to that found considering the change in direction $\theta$. Notably there is significant anticorrelation in $\phi$ for the time interval $[2\,$h$<\delta t<5\,$h], in agreement with the biased nature of the random walk. For comparison, unbiased random walk has no such anticorrelation.

\newpage
\subsection{Single cells: direction of motion and cell elongation}
\label{sec:morph}

\begin{figure}[b]
\centering
\includegraphics[width=\textwidth]{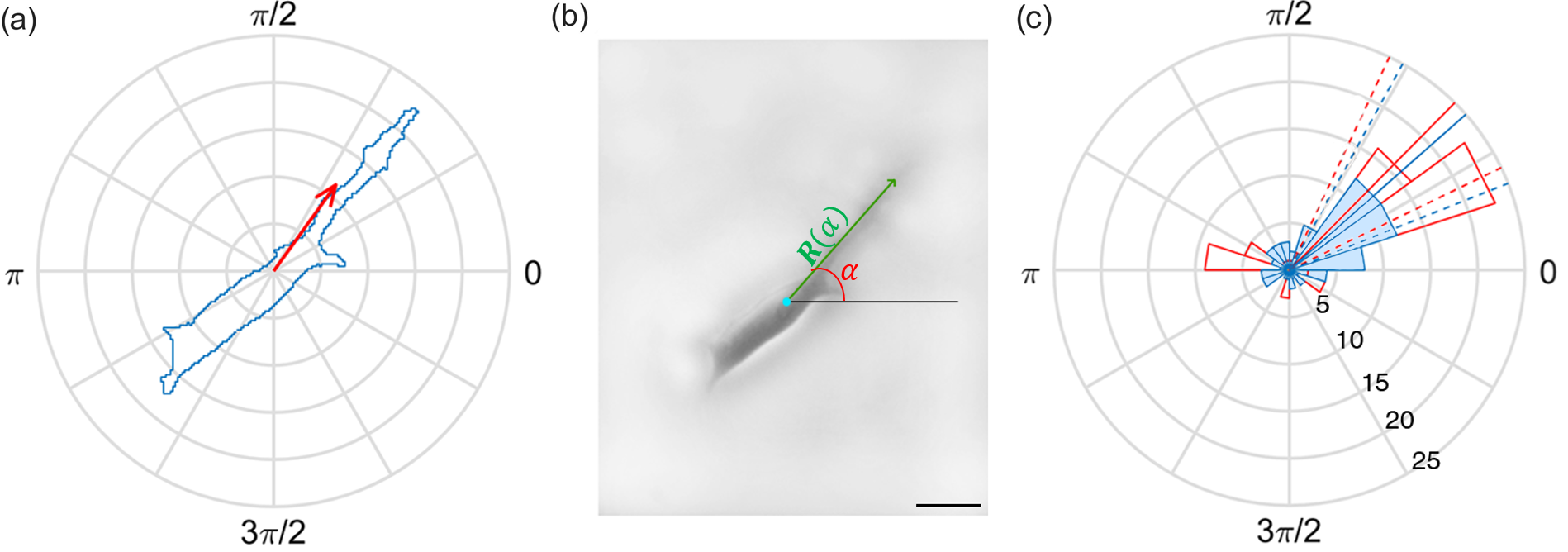} 
\caption{\label{fig:cellframe}(a) Example of the directed cell walk showing the outline of the cell (blue) and the geometric centre velocity (red arrow). (b) Microscopy image of the cell showing the geometric centre (blue dot), illustrating the definitions of the alignment angle $\alpha$ and the distance from the geometric centre to the edge of the cell $\mathbf{R}(\alpha)$. The scale bar is 30$\,\mu$m. (c) Polar histogram of the alignment angle $\alpha$ (red) and the direction of travel $\phi$ (blue, shaded) for the cell over a period of 17.5 hours, with 70 measurements in 20 bins. The mean values are shown for $\phi$ (blue, $0.72\pm0.35$) and $\alpha$ (red, $0.79\pm0.34$) with the 95\% confidence intervals for the mean shown as dashed lines.} 
\end{figure}

It is evident, from the images in Figure~\ref{fig:fb} in particular, that the direction of motion appears to be aligned with the elongation axis of the cell structure including its pseudopodia. A further example is shown in Figure~\ref{fig:cellframe}. This is unsurprising as cell branching and elongation has been shown to be involved in cell motion and directional persistence, although it has not been fully quantified \cite{LamellipodiumReview}.

To analyse quantitatively the alignment of the direction of motion and the elongation of the cell we measure the alignment angle of the cell, $\alpha$, with respect to a global reference frame. Consider $\mathbf{R}(\alpha)$, the vector from the geometric centre to the boundary of the cell and $\mathbf{R}_\mathrm{max}$ corresponding to the maximum magnitude of $\mathbf{R}$. The alignment angle $\alpha$ is defined as the angle between $\mathbf{R}_\mathrm{max}$ and the horizontal, as shown in Figure~\ref{fig:cellframe}(b). The polar histograms of $\alpha$ and the direction of travel on the plate, $\phi$, both in the same global reference frame, are shown in Figure~\ref{fig:cellframe}(c). Their mean values are $\overline{\alpha}=0.79\pm0.34$ and $\overline{\phi}=0.72\pm0.35$. The difference is insignificant as the Watson--Williams and Kuiper's tests provide no evidence to reject the null hypothesis that $\overline{\alpha}$ and $\overline{\phi}$ are from the same distribution at the 99\% confidence level. 

\begin{figure}[t]
\centering
\includegraphics[width=0.5\textwidth]{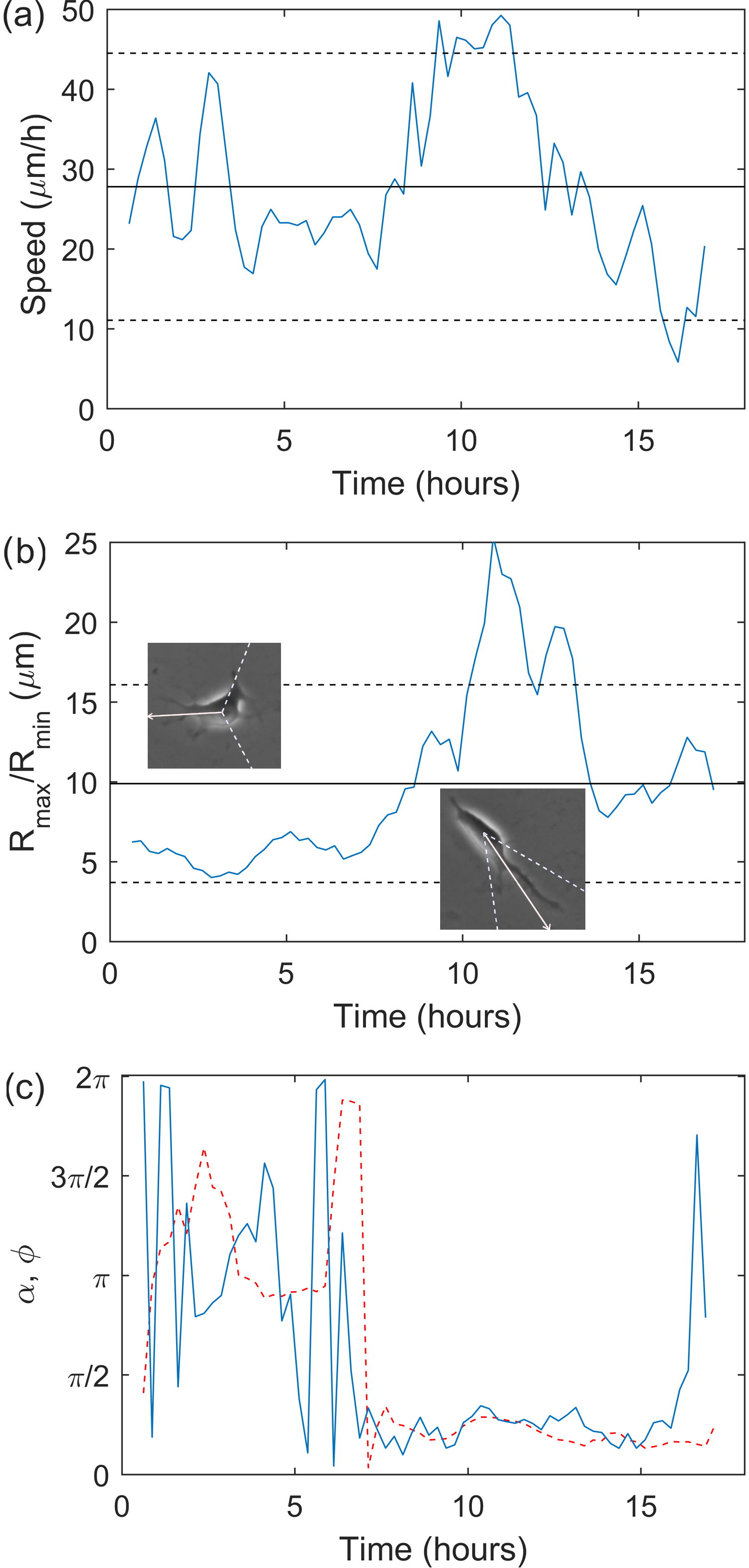} 
\caption{\label{fig:vralphaphi}Hourly moving average of (a) the cell migration speed, $v$, and (b) $R_\mathrm{max}/R_\mathrm{min}$ over time. The solid lines show the mean values of $\overline{v}=27.8\,\mu$m/h and $\overline{R_\mathrm{max}/R_\mathrm{min}}=9.98$, with dashed lines one standard deviation from the mean ($\sigma_{v}=16.7$ and $\sigma_{R_\mathrm{max}/R_\mathrm{min}}=6.2$). Insets show the cell at 4.5 and 11.5\,h with a red arrow indicating the two-hourly average direction of the velocity with white dashed lines $\pm 1$ standard deviation. (c) Hourly moving average of $\alpha$ (red, dashed) and $\theta$ (blue, solid) versus time.} 
\end{figure}

The speed of migration, $v$, and the measure of elongation of the cell, $R_\mathrm{max}/R_\mathrm{min}$,  where $R_\mathrm{max}=\mathrm{max}|\mathbf{R}|$ and $R_\mathrm{min}=\mathrm{min}|\mathbf{R}|$, are shown as functions of time in Figure~\ref{fig:vralphaphi}. The hourly moving averages of $R_\mathrm{max}/R_\mathrm{min}$ and the cell speed $v$ have a Pearson correlation coefficient of 0.53 suggesting a slight positive correlation between the elongation of the cell and its speed. Hourly moving averages of $\alpha$ and $\phi$ are shown in Figure~\ref{fig:vralphaphi}(c). This shows that directed movement is in the direction of the pseudopodia and suggests that the cell moves faster when it is more elongated. 

\newpage
\newpage
\subsection{Pairs of cells}
Wadkin \textit{et al}. \cite{me} considered the movement of cell pairs (two cells within $150\,\mu$m of each other at the start of imaging) as a whole and found that the motion of their geometric centre is approximated by an isotropic random walk for up to around 7 hours of their evolution, albeit with reduced motility compared to that of single cells \cite{me}. The diffusivity is reduced from $80\,\mu$m$^2$/h for single cells, to $60\,\mu$m$^2$/h for pairs. In this section we look in greater detail at the dynamics of pairs of hESCs, in particular the correlations between the individual motions of a pair's cells.

\begin{figure}[t]
\centering
\includegraphics[width=0.6\textwidth]{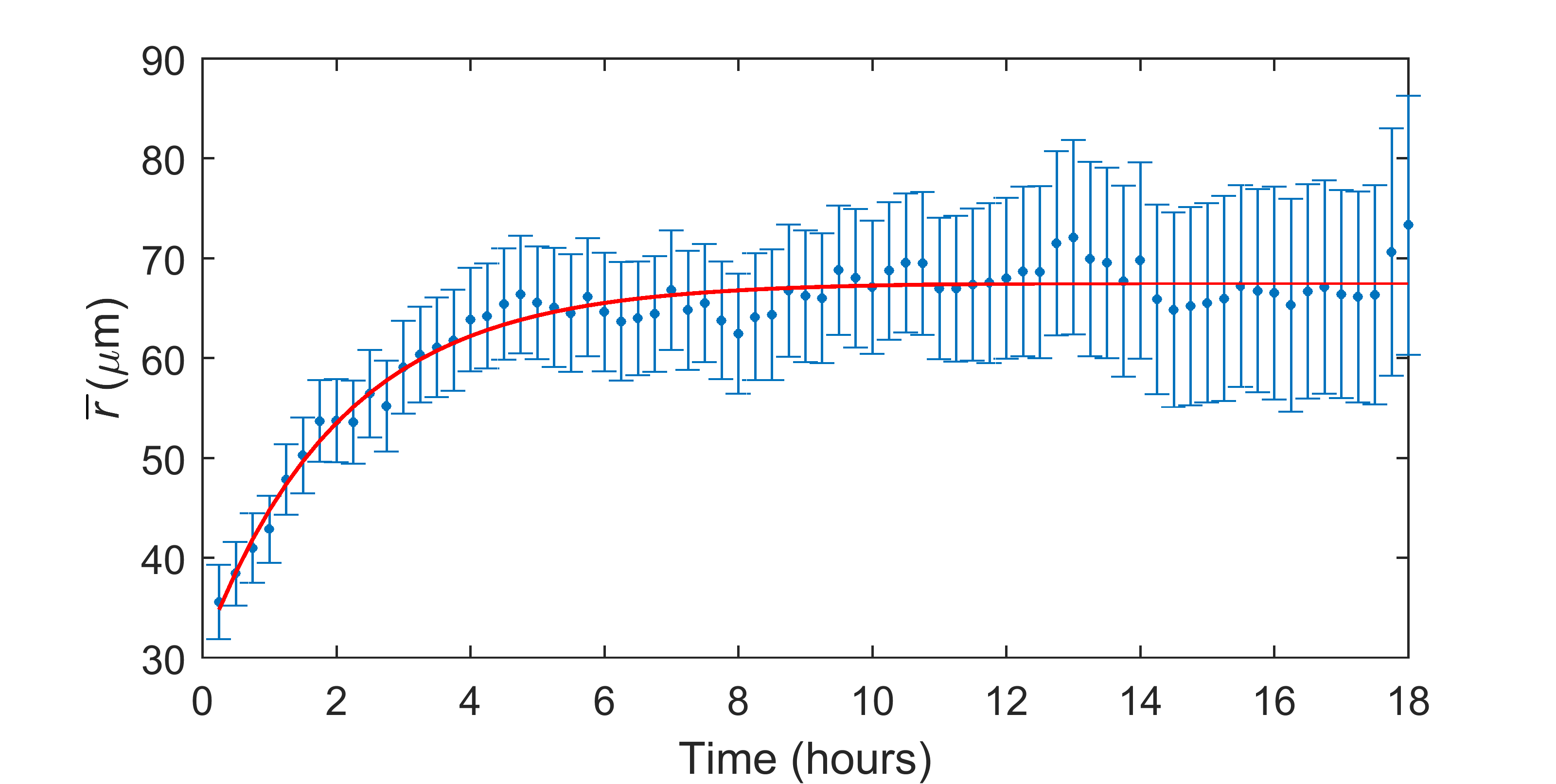} 
\caption{\label{fig:sepovertime}The mean separation, $\overline{r}$, for pairs over time with least-squares line of best fit $\overline{r}=68-35e^{-t/2}$ and $R^2=0.94$. The error bars show the standard error in the mean ($\sigma/\sqrt{N}$).} 
\end{figure}

For the cell pairs in the experiment, the mean separation at time $t$, $\overline{r(t)}=\overline{\sqrt{(\delta x(t))^2 +(\delta y(t))^2}}$, where $\delta x$ and $\delta y$ are the distances between two cells in the $x$ and $y$ directions respectively, varies with time as shown in Figure~\ref{fig:sepovertime}. By performing a least-squares fit of the functional form $\overline{r}=\mathrm{A}-\mathrm{B}e^{-t/\mathrm{C}}$, for parameters A, B and C we obtain the line $\overline{r}=(68\pm0.6)-(37\pm3) e^{-t/(2\pm0.03)}$. The asymptotic nature of $\overline{r}$ indicates an optimal separation of pairs at around 70$\,\mu$m. 

\begin{figure}[t]
\centering
\includegraphics[width=\textwidth]{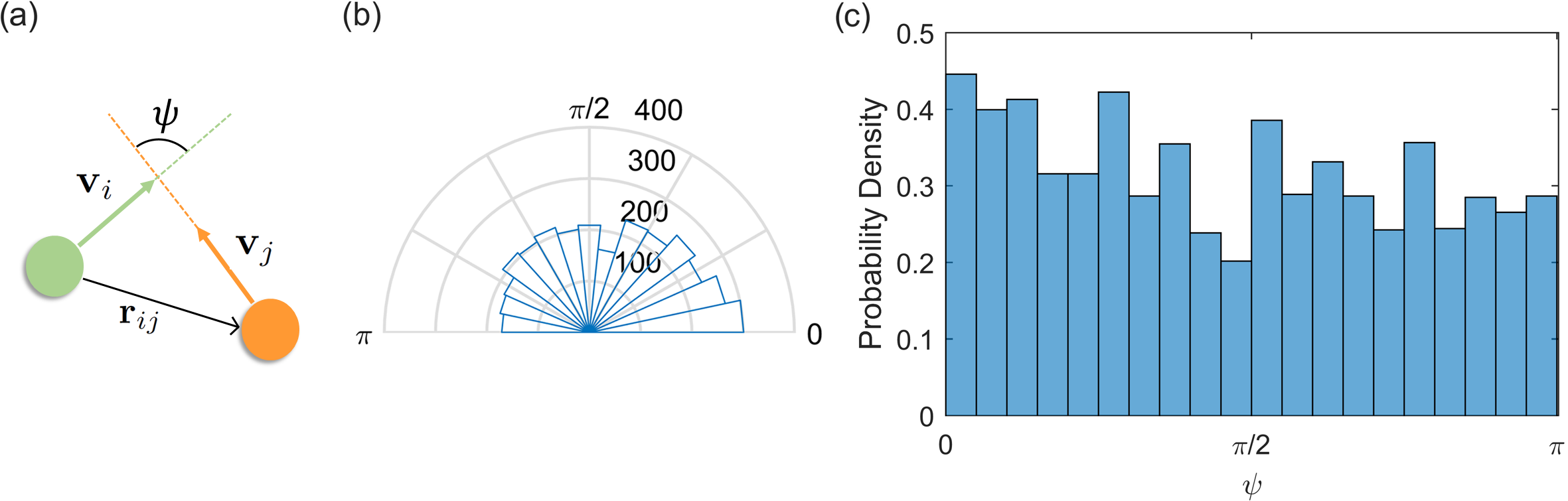} 
\caption{\label{fig:psi}(a) Green and orange dots represent a pair of cells with their corresponding velocity vectors $\mathbf{v}_i$ and $\mathbf{v}_j$ together with their connection vector $\mathbf{r}_{ij}$. The angle between the velocity vectors is marked as $\psi$. (b) Polar histogram of $\psi$ for all 50 pairs of cells. There are 15 bins and 3285 observations. (c) Corresponding linear histogram with 20 bins.} 
\end{figure}

\begin{figure}[t]
\centering
\includegraphics[width=0.9\textwidth]{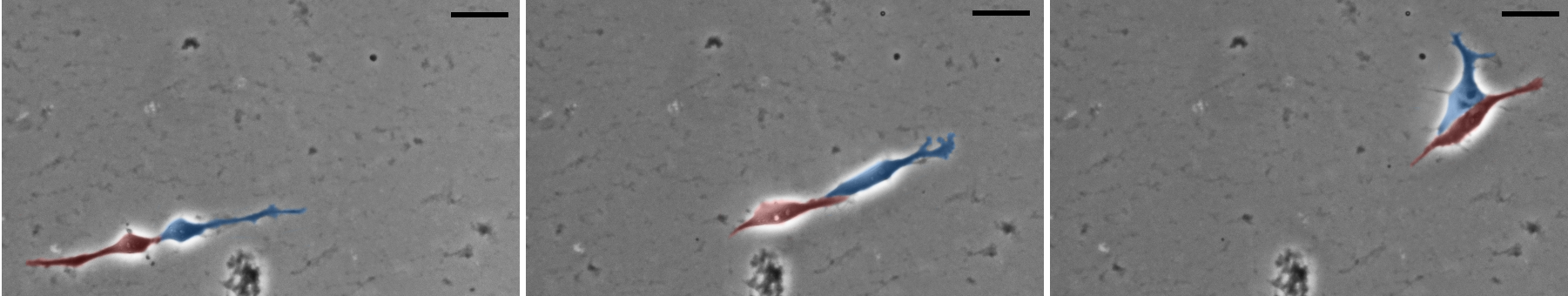} 
\caption{\label{fig:pair}Example pair moving together in the same direction. The frames are at 7\,h\,15\,min, 19\,h\,30\,min and 24\,h\,15\,min. The scale bar shows 20$\,\mu$m.} 
\end{figure}

To quantify the coordination between the movements of the two cells in a pair, we measure the smaller angle between their velocities, $0<\psi<\pi$, illustrated in Figure~\ref{fig:psi}(a). If the cells travel in the same direction on the plate, then $\psi=0$, and if they travel in opposite directions $\psi=\pi$; note that $\psi=\pi$ does not distinguish between the two cells moving exactly towards each other or exactly apart. The polar histogram of $\psi$ for all the pairs is shown in Figure~\ref{fig:psi}(b), with the corresponding linear histogram in Figure~\ref{fig:psi}(c). There is a bias in the distribution towards $\psi=0$, confirmed by the $\chi^2$ test which rejects the null hypothesis that the distribution is uniform at the 95$\,\%$ level, i.e., there is a significant preference towards pair cells moving in the same direction. Example microscopy images of a pair that move in this way are shown in Figure~\ref{fig:pair}. 

\begin{figure}[t]
\centering
\includegraphics[width=\textwidth]{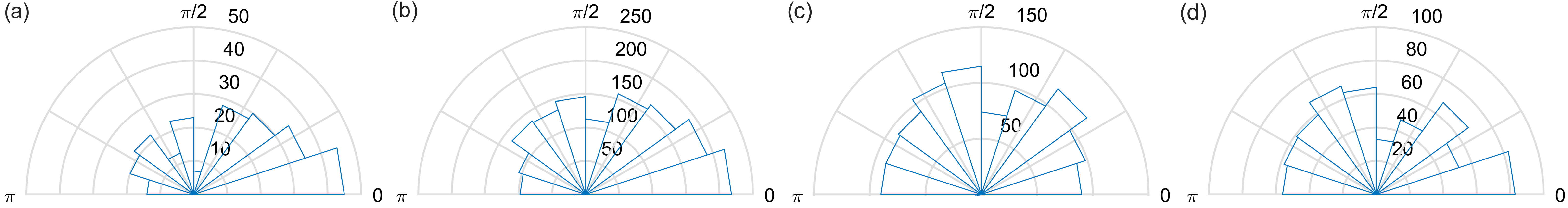} 
\caption{\label{fig:psibysep}The angle between velocity vectors, $\psi$, for separations $r$ (a) $<20\,\mu$m, (b) 20--50$\,\mu$m, (c) 50--100$\,\mu$m and (d) $>100\,\mu$m, with 20 bins and 240, 1480, 974 and 591 measurements, respectively.} 
\end{figure}

\begin{figure}[h]
\centering
\includegraphics[width=.6\textwidth]{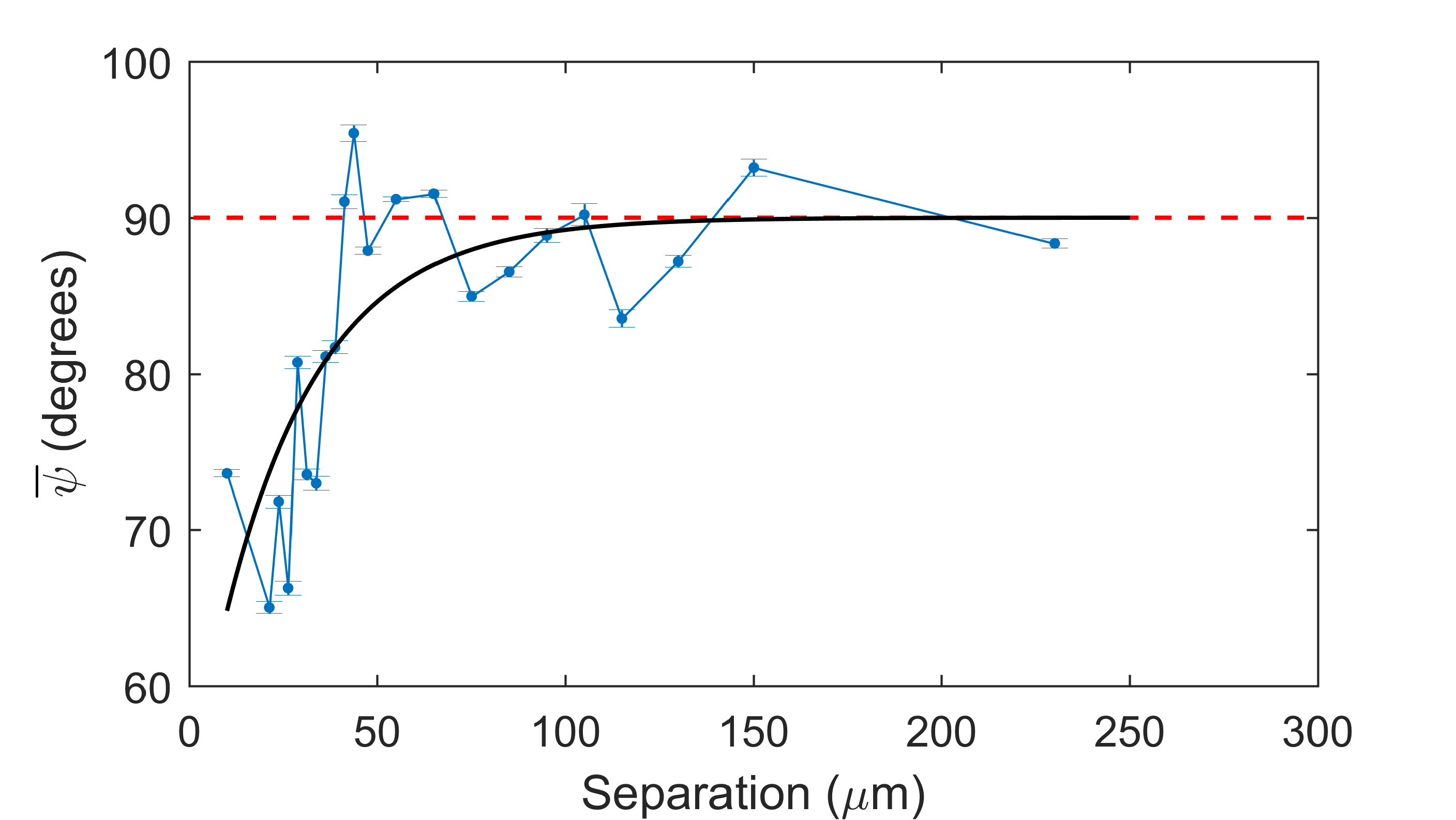} 
\caption{\label{fig:meanpsi}$\overline{\psi}$ binned according to the separation distance, $r$, between two cells. Error bars show the standard error in the mean ($\sigma/\sqrt{N}$). The red dashed line shows 90$\,\degree$, the value we would expect for uncorrelated motion. The least-squares fit (solid black line) is $\overline{\psi}=90(1-e^{-(r+r_0)/m})$, with $r_0=23.0$ and $m=25.9$ and an $R^{2}$ value of 0.6.} 
\end{figure}

Binning $\psi$ according to the separation distance, $r$, between two cells shows that this bias primarily occurs at small separations as shown in Figure~\ref{fig:psibysep}. The $\chi ^2$ test provides evidence to reject that each of the histograms in Figure~\ref{fig:psibysep} is uniform at the $95\,\%$ level. However, a measure of the skew is shown in the first moment, i.e., the arithmetic mean, $\overline{\psi}$ (as opposed to the circular mean). For a uniform distribution between 0 and $\pi$ the arithmetic mean would be $\overline{\psi}=\pi/2$ or $90 \degree$. For the $\psi$ distributions for $r<20\,\mu$m, between 20--50$\,\mu$m, between 50--100$\,\mu$m and $r>100\,\mu$m the arithmetic mean values are respectively, $\overline{\psi}=$73$\degree$, 79$\degree$, 89$\degree$ and 88$\degree$, indicating there is bias towards $\psi=0$ at smaller separations. Pearson's moment coefficient of skewness, $\gamma=E[(\psi-\overline{\psi})^3]/\sigma_{\psi}^3$, also provides a measure of the asymmetry in the distributions. For a perfectly symmetrical distribution $\gamma=0$, while for a distribution skewed towards lower values $\gamma>0$ and for skew towards higher values $\gamma<0$. For $\psi$ where $r<20\,\mu$m $\gamma=0.40$, for $20<r<50\,\mu$m $\gamma=0.24$, for $50<r<100\,\mu$m $\gamma=0.04$ and for $r>100\,\mu$m $\gamma=-0.02$, showing reducing skewness towards $\psi=0$. The Kolmogorov-Smirnov test provides no evidence to reject the null hypothesis that the distributions for $r<20\,\mu$m and  $20<r<50\,\mu$m are the same. Similarly for  $50<r<100\,\mu$m and  $r>100\,\mu$m. However the test rejects the null hypothesis that the two smaller separation distributions are the same as the two larger separation distributions. Calculating $\overline{\psi}$ with separations binned more frequently shows the length at which the movement is correlated. By performing a least-squares fit of the form $\overline{\psi}=90(1-e^{-(r+r_0)/m})$, for parameters $r_0$ and $m$, we obtain the line $\overline{\psi}=90(1-e^{-(r+23.0)/25.9)})$ with an $R^{2}$ value of 0.6, shown in Figure~\ref{fig:meanpsi}. The characteristic length of the decay is therefore 26$\,\mu$m, a typical cell diameter, suggesting the pairs only exhibit correlated motion while they are in contact. 

We also analysed the motion of the cells in the pair via the pair correlation function.  This function was not found to be sensitive to the correlations between the cell motions, and was unable to distinguish the cell motions from from IRWs. This analysis is presented in the Appendix.

\newpage

\newpage
\section{Discussion}

\begin{figure}[b]
\centering
\includegraphics[width=.58\textwidth]{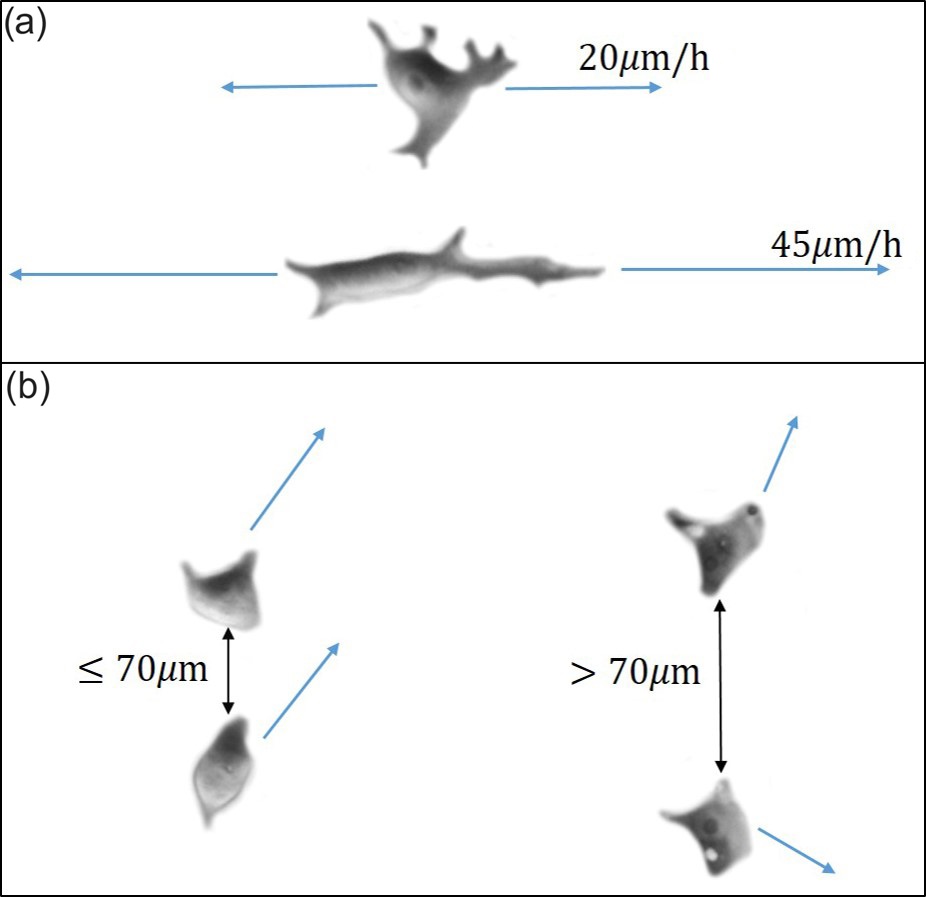} 
\caption{\label{fig:schematic}(a) Single hESCs preferentially move along their elongation axis, at speed higher for a stronger elongation. (b) Cells separated by $70\,\mu$m or less move in a coordinated manner, whereas a wider separation implies independent biased random walk \cite{me}.}
\end{figure}

In culture, hESCs are anchorage-dependent: they adhere to the surface and sense external cues by extending lamellipodia and filopodia, referred to in a general way as pseudopodia. For directed movement in response to external factors, cells acquire a defined front-rear polarity extending a protrusive structure at the leading edge before subsequently moving the cell body, and retracting the trailing edge \cite{migration}.  The integration of negative and positive chemical feedback loops accounts for the oscillatory behaviour of pseudopodia, i.e. cycles of protrusion and retraction which result in cell movement \cite{eukaryoticreview}. Observations of single cell movement in two-dimensions cultures, in the absence of external cues, indicate a production of pseudopodia structures in random directions, a behaviour observed in other cell types \cite{Reig1999}.

Our results are summarised in Figure~\ref{fig:schematic}. The relative angle of movement, $\theta$, characterises the dynamics of random walks further to the mean-square displacement \cite{DirectionalChange}. Our results show that isolated single cells migrate in an unusual uni-directional walk, moving backwards and forwards along a preferred local axis, with cells becoming more persistent over time. Hence, the longest lived isolated cells show the strongest directional persistence. Broadly, there are a wide range of example cells that exhibit a preferential turning angle; those that can be modelled as a correlated random walk as previously discussed, e.g., \cite{CRWamoeboid, CRWepithelial, CRWfibroblasts}. There are also examples of a biomodal preference for turning angle, similar to the one we see for single hESCs \cite{BimodalTurns1, BimodalTurns2}. The bias in the walk is further shown in the temporal correlation in both the change in direction, and the direction of movement with a correlation time of around 0.8$\,$h. The microscopy images in Figure~\ref{fig:fb} show the elongated morphology of the single cells, with movement in the direction of the leading pseudopodia, leading to this motion along a local axis.  

These single cells demonstrate random migratory patterns, travel large distances and do not result in colony formation. Isolated cells seeded at low density display directional migration towards neighbours \cite{Phadnis15}. Perhaps in the absence of neighbours, as in this experiment, the cells employ the uni-directional walk along the local axis in an attempt to locate neighbours. Our quantitative analysis of a directed cell trajectory confirms the axis of cell motion is aligned with the elongation axis of the cell. Increased elongation is also linked to increased speed, corresponding to previous results suggesting that persistence in direction of motion is linked to increased speed as a universal rule for all types of cells \cite{speedpersist}. 

An understanding of the migration of single hESCs is integral to colony growth at low-density platings. Their directed, super-diffusive migration can facilitate colony expansion at low-density platings by the finding and joining neighbours, however this re-aggregration is undesirable in experiments which require colonies originating from a single cell to achieve a homogenous clonal population \cite{Li, Andrews}. 

For pairs of hESCs their separation over time increases exponentially before approaching an asymptote at a distance of $70\,\mu$m. This shows that, on average, $70\,\mu$m is the optimal separation for pairs of cells. There is a preference for the cells to move in the same direction as each other on the plate at small separations ($<\,70\,\mu$m). At these small separations it can be seen from the microscopy imaging that the cells are physically connected by their pseudopodia, as in Figure~\ref{fig:pair}. This coordinated movement could be due to an external stimulus, but the connection of the cell bodies facilitates this motion. At separations greater than $\approx 70\,\mu$m the motion of each cell in a pair appears uncorrelated. Often there is still a connection between the cell bodies at these distances, but the cells move in independent directions whilst maintaining the connection, and as an isotropic random walk when considered as a whole entity \cite{me}. Neighbouring cells are integral to colony formation as cell survival and cell divisions are highly correlated with the number of neighbouring cells \cite{Phadnis15}. 

Another ramification would be an exploration of the effects of stem cell markers, such as NANOG, OCT and KLF, on cell migration. These factors have been shown to affect the migration, invasion and colony formation of various cancer stem cells \cite{pmid22945654, pmid25821200, pmid21242971}. Effects of pluripotency markers on the migration and motility of single hESCs have not been explored. hESCs with NANOG overexpression form colonies efficiently even at very low seeding densities. Cell motility and colony formation affected by stem cell markers are subjects of our future work. 

Further experiments need to verify the robustness of these results under different culture conditions. This additional information on low density plated cells will assist in the development of agent-based models, combining the motion of diffusive and super-diffusive cells with their biological states and cell-cell interactions. 

\ack
We acknowledge financial support from Newcastle University and European Community (IMI-STEMBANCC, IMI-EBISC, ERC \#614620 and NC3R NC/CO16206/1) and are grateful to the School of Mathematics, Statistics and Physics of Newcastle University (Prof. R. Henderson) for providing partial financial support. AS acknowledges partial financial support of the Leverhulme Trust (Grant RPG-2014-427).

\appendix
\label{appendix}
\section*{Appendix}
\setcounter{section}{1}

The pair correlation measures to what extent the direction of motion of each cell is correlated to that of the other \cite{corr}. To compute the correlation in the motion of the paired cells, we calculated the projections of the directions of the individual velocities of each cell, at each time frame $t_1, t_2, ..., t_\mathrm{N}$, $\mathbf{v}_1(t_\mathrm{k})$ and $\mathbf{v}_2(t_\mathrm{k})$, onto the vector $\mathbf{r}_{12}(t_\mathrm{k})$ joining them at each time step, as illustrated in Fig.~\ref{fig:psi}. The correlation function for one pair is defined as,

\begin{equation}\label{eq:corr}
C(r) = \frac{1}{2} \Bigg[ \frac{ \sum_{k=1}^{N} \mathbf{\hat{v}}_1 \cdot \mathbf{\hat{r}}_{12} \ \delta(r-r_{12})}{ \sum_{k=1}^{N} \delta(r-r_{12})} + \frac{ \sum_{k=1}^{N} \mathbf{\hat{v}}_2 \cdot \mathbf{\hat{r}}_{21} \ \delta(r-r_{21})}{ \sum_{k=1}^{N} \delta(r-r_{21})} \Bigg]
\end{equation}

\noindent where circumflex denotes a unit vector, $r_{12}=|\mathbf{r}_{12}|$, $\delta(r-r_{12})$=1 if $r<r_{12}<r+\delta r$ and zero otherwise, and $\delta r$ is the width of a bin. A positive correlation indicates that the cells tend to approach one another, whereas $C(r)<0$ indicates that they systematically move apart. The cells in pairs with $C(r)\approx0$ move with little or no coordination. 

\begin{figure}[t]
\centering
\includegraphics[width=\textwidth]{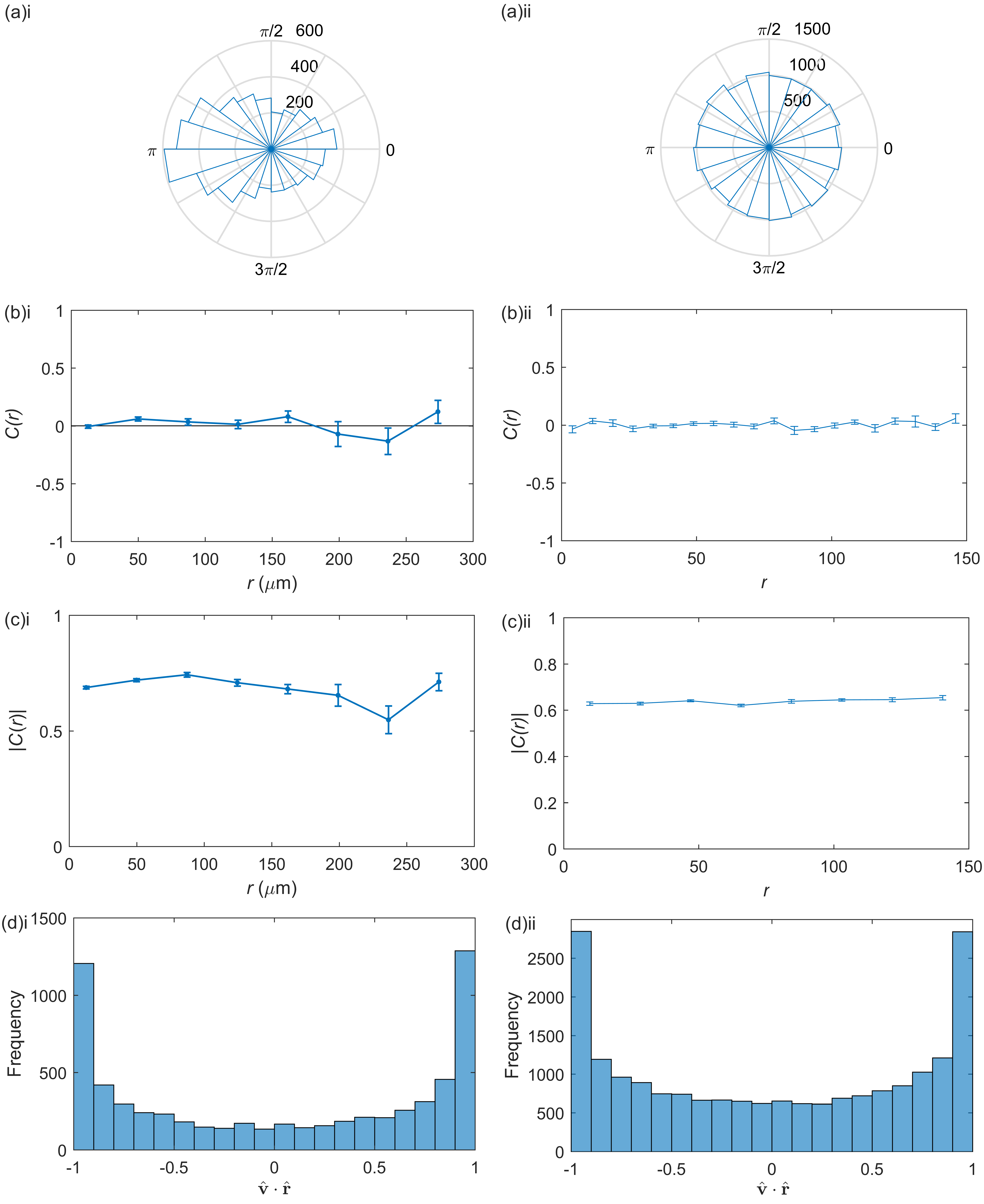} 
\caption{\label{fig:compare}(a) The angle of movement relative to the last, $\theta$, (b) the correlation $C(r)$, (c) the absolute correlation $|C(r)|$ and (d) $\mathbf{\hat{v}} \cdot \mathbf{\hat{r}}$ for i) the experimental pairs and ii) a simulated IRW for two cells. $|C(r)|$ is theoretically constant at $2/\pi$ for an uncorrelated IRW pair.} 
\end{figure}

The pair correlation for all 50 pairs considered together is approximately zero due to the averaging of positive and negative correlations, see Figure~\ref{fig:compare}. However, we can assess the average degree of correlation (positive or negative) by considering the magnitude of the correlation, $|C(r)|$. The absolute value of the correlation for all pairs, calculated by taking $|\mathbf{\hat{v}}_i \cdot \mathbf{\hat{r}}_{ij}|$ in Eq.~\ref{eq:corr} and is within errors to the equivalent for a random isotropic walk for both cells in the pair. A comparison of $\theta$ (the angle of movement for each individual cell), $C(r)$ and $|C(r)|$ for the experimental data and for a simulated IRW for both cells is shown in Figure~\ref{fig:compare}. For an IRW with no correlation between cells in a pair, the expected value of $|C(r)|$ is $2/\pi$, resulting from $E[|\cos(\theta)|]=2/\pi$.\\ 

\newpage

\newpage

\newpage
\section*{References}
\bibliographystyle{unsrt}
\bibliography{mybib}

\end{document}